# Manganese oxide nanosheets and a 2D hybrid of graphene-manganese oxide nanosheets synthesized by liquid-phase exfoliation

Joao Coeiho, Beatriz Mendoza-Sanchez ᵛ Henrik Pettersson, Anuj Pokle, Eva K. McGuire, Edmund Long, Lorcan McKeon, Alan P. Bell, and Valeria Nicolosi

*Trinity College Dublin, Schools of Chemistry and Physics and CRANN, Dublin 2, Ireland*

E-mail: beatrix.mendoza@gmail.com

"These authors contributed equally to this work

Abstract

Manganese oxide nanosheets were synthesized using liquid-phase exfoliation that achieved suspensions in isopropanol with concentrations of up to 0.45 mg ml$^{-1}$. A study of solubility parameters showed that the exfoliation was optimum in N,N-Dimethylformamide followed by isopropanol and diethylene glycol. Isopropanol was the solvent of choice due to its environmentally friendly nature and ease of use for further processing. For the first time, a hybrid of graphene and manganese oxide nanosheets was synthesized using a single-step co-exfoliation process. The 2D hybrid was synthesized in isopropanol suspensions with concentrations of up to 0.5 mg ml$^{-1}$ and demonstrated stability against re-aggregation for up to 6 months. The co-exfoliation was found to be a energetically favorable process in which both solutes, graphene and manganese oxide nanosheets, exfoliate with an improved yield as compared to the single-solute exfoliation procedure. This work demonstrates the remarkable versatility of liquid-phase exfoliation *To whom correspondence should be addressed with respect to the synthesis of hybrids with tailored properties, and it provides proof- of-concept ground work for further future investigation and exploitation of hybrids made of two or more 2D nanomaterials that have key complementary properties for various technological applications.

# 1 Introduction

The role of two-dimensionality on bringing up unique properties of a material, not present in its bulk counterpart, was first revealed in the case of graphene. Graphene showed exceptional electronic, optical and mechanical properties due to the confinement of charge and heat in a plane.[1,2] Shortly thereafter, the research community realized that graphene was only the first of a series of 2D nanomaterials with unusual physical-chemical properties.[2-4] Due to the large diversity of existent layered compounds, from which 2D nanomaterials are derived, an almost infinite variety of 2D layered compounds can be produced.[5] Various families of layered compounds have been identified, and they include boron nitride, transition metal di- and tri-chalcogenides,[6,7] layered metal oxides,[8,9] metal halides, transition metal carbides[10] and nitrides, and layered double hydroxides.[8]

Inorganic 2D nanomaterials are highly attractive because they exhibit unusual properties absent in their bulk counterparts. As in the case of graphene, quantum confinement in two dimensions leads to a significant modification of electronic structure, which induces novel physical phenomena.[2] For instance, the electronic band structure of $MoS_2$ is altered as the single-layer limit is approached.[4,11] 2D nanomaterials have an intrinsic high surface area and flexibility, and some of them exhibit unusually high mechanical strength, high electrical and thermal conductivities, and high dielectric constants.[2] Therefore, 2D nanomaterials are expected to improve current device technology including electronic, ferromagnetic, magneto- optical, electrochemical, and photoresponsive nanodevices.[2]

Further interest on 2D nanomaterials resides in the possibility to combine them in rationally designed functional materials with enhanced properties for various applications includ ing photocatalysis, biosensing, devices using superparamagentic films and devices generating photocurrent and photoluminiscence.[29] 2D nanomaterials have been used as building blocks of hierarchically organized nanostructures manufactured by various methods: layer-by-layer deposition (multilayers, superlattices),[12,13] Langmuir-Blodgett deposition (multilayers, su- perlatttices),[12,13] flocculation (ions interstratified with 2D nanomaterials).[14-16] Moreover, the multilayers can then be further engineered[8,9,12] for instance introducing pores using tem- plating techniques[17] or freeze-drying[18] and forming microspheres by spray-drying.[19] Most of the methods mentioned yield satisfactory results at the laboratory scale but result unpractical for commercial applications due to their time-consuming multi-step procedures. It is clear that the versatility of 2D nanomaterials can be exploited only if cost-effective and scalable methods are developed.

In the case of energy storage applications, 2D transition metal oxides (TMOs) are of high relevance[20] due to their high surface area available for faradaic charge storage and short transport paths for electrons and ions.[4] Amongst several transition metal oxides, manganese oxide is a material of choice due to its low toxicity, low cost and natural abundance.[21,22] It has been shown that manganese oxide uses only a nanometer-thick portion of its surface to store charge. Therefore a high surface area 2D nanosheet with nanometer thickness results ideal for enhanced charge storage.[23] Moreover, manganese oxide with a high surface area is relevant not only for energy storage applications but also in the catalysis of several reactions,[24,25] as adsorbent in waste-water treatment[26,27] and for use in electrochromic devices.[28]

Two-dimensional manganese oxide nanosheets have been synthesized by various methods: chemical exfoliation of layered $K_{0.45}MnO_2$[29] and Birnessite-type manganese oxide[30] conducted via a multistep procedure involving intercalation first of protons and then of bulky tetraalkylammonium ions; reduction of $KMnO_4$ with surfactants;[31,32] oxidation of $Mn(NO_3)_2$ or $MnCl_2$ with $H_2O_2$ in the presence of tetramethylammonium hydroxide[33,34] (20-40 nm sheets); reduction of $KMnO_4$ using graphene oxide as reductant and template;[35] synthesis of $MnO_2$ nanosheets as part of a ternary composite where the $MnO_2$ nanosheets are grown on top of gold or a form of carbon.[36-38] Although these methods represent clear advances on the manufacturing of $MnO_2$ nanosheets, various areas of opportunity remain. Current synthesis procedures often involve multi-step, time-consuming and poorly reproducible procedures (wet chemistry with or without a templating agent, electrodeposition, hydrothermal treatments and annealing) that may use toxic and expensive chemicals and treatments at high temperatures requiring inert atmospheres. Overall these methods result expensive and do not offer scope for scalability.

In this study, we present a facile, reproducible and scalable method for synthesizing $MnO_2$ nanosheets with lateral dimensions of 20-40 nm and thickness of 3.2 nm using a combination of a known room-temperature wet chemistry method and liquid-phase exfoliation. Liquid- phase exfoliation is a well established, scalable and cost-effective method for producing stable suspensions of a variety of 2D nanomaterials[6-39] that can in turn be easily processed using other equally scalable methods for producing flexible films such as spray deposition[40] and ink-jet printing. Moreover, for the first time, a hybrid of manganese oxide nanosheets and graphene was produced using a facile single-step liquid-phase co-exfoliation method. The hybrid strategically combines the high electrical conductivity properties of graphene and the pseudocapacitive properties of manganese oxide nanosheets in a 2D hybrid of superior energy storage properties as described extensively elsewhere (work in preparation). This study demonstrates the remarkable versatility of liquid-phase exfoliation in the production of functional hybrids and it provides the basis for future synthesis of hybrids of two or more 2D nanomaterials that have properties tailored to numerous applications.

# 2 Experimental methods

Materials. Manganese nitrate tetrahydrate ($Mn(NO_3)_2\cdot H_2O$, purity > 97.0 %), graphite flakes, potassium permanganate ($KMnO_4$, BioUltra, purity > 99.0 %), poly(ethylene glycol)- block-poly(propylene glycol)-block-poly(ethylene glycol) (PEG-PPG-PEG, average Mw =

5,800 g mol$^{-1}$) were supplied by Sigma Aldrich; all solvents N,N-Dimethylformamide (DMF), isopropyl alcohol (IPA), diethylene glycol, ethanol, benzyl alcohol, bromonaphthalene, dibro- momethane, N-methyl-2-pyrrolidone (NMP), cyclohexanone, Benzyl Benzoate, cyclohexane, acetone and formamide were supplied by Sigma Aldrich and used as received for liquid-phase exfoliation without further purification; deionized water (Dl-water, 10 MO-cm) was used for all the synthesis protocols.

Synthesis of manganese oxide flower-like nanostructures (MOFN). Manganese oxide was synthesized following the method described by Hao J. et al,[41] Mn(NO$_3$)$_2$-H$_2$O (1.67 g, 6.66 mmol) was dissolved in Dl-water (100 ml), the triblock copolymer PEG-PPG-PEG (0.5 g) was then added. A second aqueous solution of KMnO$_4$ (0.1 M, 100 ml, 0.01 mol) was prepared. The first solution was heated up and maintained at a given temperature (5 °C, 25 °C, 45 °C, 60 °C, 75 °C and 90 °C) using a hot plate and was kept under vigorous stirring while the second solution was added drop by drop. The obtained brownish precipitate was washed thoroughly, first with DI-water and then with ethanol; it was then vacuum filtrated and dried in an oven at 50 °C overnight.

Liquid-phase exfoliation I. Various MOFN samples were prepared at the various temperatures specified above (300 mg) and each sample was mixed with IPA (30 ml). The mixture was processed in an ultrasonic bath (37 kHz, 198 W = 0.6P where P is the nominal power of 330 W, 3 hours), followed by centrifugation (5000 rpm/RCF = 4662, 3 hours) and collection of the supernatant (top 70 %). The quality of the obtained dispersions was assessed by UV-Vis spectroscopy and TEM. A MOFN synthesis temperature of T = 45 °C was selected as standard.

Liquid-phase exfoliation II: solvent selection experiments. The as-prepared MOFN (60 mg) was mixed with water (30 ml) and each of the solvents (30 ml) listed in the Materials section. Each mixture was processed in an ultrasonic bath (37 kHz, 198 W, 3 hours), followed by centrifugation (5000 rpm/RCF = 4662, 3 hours) and collection of the supernatant (top 70 %). The quality of the obtained dispersions was assessed by UV-Vis spectroscopy.

Liquid-phase exfoliation III: optimization of processing parameters. The as-prepared MOFN was mixed with IPA (30 ml) and processed in an ultrasonic bath following the experimental conditions described in Table S3. The dispersions were then centrifuged (5,000 rpm/ RCF = 4,662, 3 hours) and the supernatants were collected (top 70 %). The quality of the obtained dispersions was assessed by UV-Vis spectroscopy and TEM. The optimized exfoliation parameters were as follows: initial MOFN/IPA mixture concentration of 20 mg ml$^{-1}$, ultrasonication power of 99 W, ultrasonication frequency of 80 kHz, and processing time of 9 hours. The final dispersions consisted of manganese oxide nanosheets (MON).

Synthesis of graphene/manganese oxide nanosheest hybrid (GMOH) dispersion. The as-prepared MOFN (300 mg) and as-purchased graphite flakes (300 mg) were mixed with IPA (30 ml). The mixture was processed in a ultrasonic bath (37 kHz, 198 W, for 3, 9 and 18 hours) followed by centrifugation (5000 rpm/RCF = 4695, 3 hours) and collection of the supernatant (top 70 %). To determine concentration, in a second experiment, a GMOH/IPA dispersion (V= 100 ml) was vacuum filtrated (using an alumina membrane, pore size of 0.02 m) to obtain the solids that were then dried (vacuum oven at 100 °C for 5 hours) and the mass m determined using a microbalance. The concentration C was determined as C = m/V.

Dispersions for co-exfoliation studies. GMOH/IPA dispersions were prepared as described above (for 3, 6, 9, 15 and 18 hours). In separate mixtures, as-purchased graphite flakes (300 mg) and MOFN (300 mg) were mixed with IPA (30 ml). The mixtures were processed in a ultrasonic bath (37 kHz, 198 W, for 3, 6, 9, 15 and 18 hours) followed by centrifugation (5000 rpm/RCF = 4695, 3 hours) and collection of the supernatant (top 70 %). Dispersion concentrations were obtained using vacuum filtration (GMOH) and UV-Vis spectroscopy and the *Beer-Lambert* law (MON and graphene).

Equipment and characterization techniques. Transmission electron microscopy

(TEM) images were obtained using a *FEI-Titan* operated at 300 keV and equipped with EDAX energy-dispersive x-ray (EDX) spectrometer and a Gatan Imaging Filter (GIF); Scanning Electron Microscopy (SEM) images were obtained in a *Zeiss Ultra Plus microscope* operated at 5 keV; Helium Ion Microscopy (HIM) images were obtained using a *Zeiss Orion microscope* operated at an acceleration voltage of 30 kV and current of 1.2 pA. A charge neutralization system was applied using an electron flood gun; X-ray diffraction (XRD) was performed in a fully automated *Bruker D5000 powder difractometer* equipped with a monochromatic Cu Ka radiation source (A = 0.15406 nm) and a secondary monochromator. XRD patterns were collected between 10 ° < $2\theta$ < 80 °, with a step size of $2\theta$ = 0.05 ° and a count time of 12 s/step. The samples were supported on monocrystalline silicon; Raman (RS) spectra were recorded at room temperature using a *Witec Alpha 300* system with a laser excitation wavelength of 532 nm. A laser power of 0.4 mW was used with a 20x objective lens. Spectra were acquired for each sample by averaging 10 distinct spectra, each with an acquisition time of 30 s; Ultrasonication was performed in a Elmasonic P120H (Fisher FB11207) sonic bath; Centrifugation was performed in a Thermoscientifc Heraeus Multifuge X1 centrifuge; UV-Vis absorption measurements were obtained using a Varian Cary 6000i UV-Vis-NIR spectrometer and two identical Helmma Analytics 6030-UV (z626902) quartz cuvettes (10 mm light path). Most absorption spectra were recorded in a A range of 200 nm to 900 nm; Contact angles were measured using a *FTA 125 video-based* contact angle and surface tension meter; films for the measurement of contact angles were manufactured by spray deposition using a *USI Prism Ultracoat 300* spray deposition equipment; the weight of deposited films was measured using a *Sartorius Ultramicrobalance MSE-2.7S000-DF* with a 0.0001 mg readability; Thermogravimetric analysis (TGA) measurements were performed with a *Perkin Elmer Pyris 1* thermogravimetric analyzer in air at a flow rate of 20 ml min$^{-1}$ and heating rate of 10 °C min$^{-1}$ from room temperature to 900 °C. A mass of about 7 mg was used for each test; X-ray photoelectron spectroscopy (XPS) was performed in an ion pumped *VG Microtech CLAM 4 MCD* equipment using a 200 W unmonochromated Mg X-ray exci tation source (1253.6 eV), with samples supported on silicon substrates. The analyzer was operated at a constant pass energy of 100 eV for wide scans and 20 eV for detailed scans. The XPS spectra was analyzed and fitted using *CasaXPS* software and spectra calibration was done fixing the position of the $C_{1s}$ = 284.9 eV as reference. A mixture of Gaussian (70 %) and Lorentzian (30 %) functions were used for curve fitting.

## 3 Results and Discussion

### 3.1 Synthesis and characterization of manganese oxide flower-like nanostructures

Manganese oxide was synthesized following a typical co-precipitation synthetic method.[40-42] As shown in Figures 1a-1c, the as-synthesized material had a characteristic flower-like and mesoporous morphology with protruding petal-like nanostructures. Hence, the material is called manganese oxide flower-like nanostructures (MOFN). The broadness of the peaks in the XRD diffraction patterns in Figure 2 showed the poor crystallinity of the as-synthesized MOFN. In consequence, a definitive phase identification was not possible. Nevertheless, the peaks at $2\theta = 36.9°$ and $2\theta = 66.1°$ could be associated to diffraction from the planes (400) and (002), respectively, of tetragonal a-$MnO_2$·$3H_2O$ (JCPDS 44-0140).[40-42] A broad diffraction peak at low angles ($2\theta < 14°$) has been associated in the literature with a degree of organization of $MnO_6$ octahedra as a precursor of a-$MnO_2$, also defined as 1D hollandite (a phase with a 2 x 2 tunnel structure).[43-45] The peak at $2\theta = 24.3°$ (indicated with an asterisk in Figure 2 corresponds to a secondary phase that could not be identified.[42]

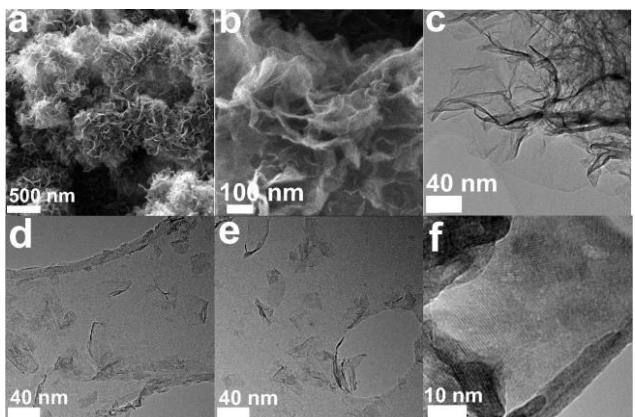

Figure 1: Micrographs of (a) MOFN (HIM), (b) close-up of MOFN (HIM), (c) close up of MOFN showing the presence of petal-like nanostructures (TEM), (d), (e) manganese oxide nanosheets (TEM) with typical dimensions of 20-40 nm, and (f) HRTEM of a manganese oxide nanosheet.

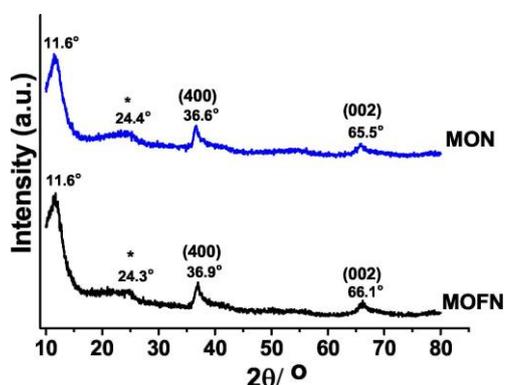

Figure 2: XRD patterns of MOFN and MON where (hkl) labels corresponds to planes assigned to the a-$MnO_2 \cdot 3H_2O$ phase. The peak labeled with * correspond to an unidentified phase.

## 3.2 Synthesis by liquid-phase exfoliation and characterization of manganese oxide nanosheets

Liquid-phase exfoliation of MOFN was attempted using isopropanol (IPA). This solvent was preferred over others due low cost, low toxicity and low boiling point, all of which which facilitates further processing. In a series of experiments, MOFN was synthesized at various temperatures (5 °C, 25 °C, 45 °C, 60 °C, 75 °C and 90 °C) and processed for exfoliation using the conditions detailed in the subsection *'Experimental*

*methods/ Liquid-*

*phase exfoliation I'*. As shown in Figure S1 (Supporting Information (SI)), the morphology of the obtained material varied with temperature having species with undefined morphology (at 5 °C), nanosheets (at 45 °C and 90 °C), partially exfoliated material with a degree of folding (at 25 °C, 60 °C), and a mix of nanosheets and partially exfoliated material (at 75 °C). Due to their expected higher surface area, flat nanosheets were preferred over folded nanosheets and a MOFN synthesis temperature of 45 °C was selected for all further experimentation. For each synthesis cited above, centrifugation speeds of 1,500 rpm, 3,000 rpm and 5,000 rpm were tested. Centrifugation speeds below 5,000 rpm resulted in dispersions containing a high fraction of partially exfoliated material. Therefore, a centrifugation speed of 5,000 rpm was selected as optimum parameter.

The manganese oxide nanosheets (MON) obtained at 45 °C synthesis temperature and 5,000 rpm centrifugation speed are shown in Figures 1d and 1e with typical lateral dimensions of 20 to 40 nm. The thickness of the MON was investigated using a "log-ratio technique" (t/A-maps technique further detailed in the SI section S7, Figure S6) and was determined as $3.2 \pm 1.2$ nm. As shown in the XRD pattern for MON in Figure 2, the exfoliation procedure did not alter the structure of the material and MON was of the same poorly crystalline nature as the starting material. As described above, the peaks at $2\theta = 36.6°$ and $2\theta = 65.5°$ were associated to diffraction from the planes (400) and (002), respectively, of tetragonal a-$MnO_2$-$3H_2O$. Figure 1 shows a HRTEM image of a MON. Although, some crystalline domains could be distinguished, and were likely produced as a result of exposure to the electron beam, the MON was mainly amorphous and acquisition of diffraction patterns and/or elucidation of a defined crystal structure by TEM techniques was not possible.

Next, we address the question of whether IPA was, in terms of yield, actually the most suitable solvent for exfoliating MOFN.

3.2.1 Solvent selection

A set of solvents was tested for exfoliation of the starting material MOFN. As detailed in the supporting information (SI section S2) and according to the solubility theory, a suitable solvent for exfoliation fulfills at least one of three criteria: (1) the surface energy of the solvent is similar to the surface energy of the solute, (2) the Hildebrand parameter of the solvent matches the Hildebrand parameter of the solute, and (3) the three Hansen solubility parameters of the solvent match the three Hansen solubility parameters of the solute. Here, the solute is the MOFN.

Under identical experimental conditions, a set of solvents with known surface tension, Hildebrand, and Hansen solubility parameters (see the subsection '*Experimental methods/Liquid- phase exfoliation II*') were tested regarding its ability to exfoliate MOFN. The solvents and corresponding surface tension and solubility parameters are listed in the Table S1. The quality of the obtained dispersions was assessed by UV-Vis spectroscopy. According to the *Beer-Lambert* law, the final concentration of dispersed material is proportional to its absorption (A/l = $aC$ where *A* is absorbance, *l* is the light path length, *a* is the extinction coefficient and *C* is the dispersion concentration).

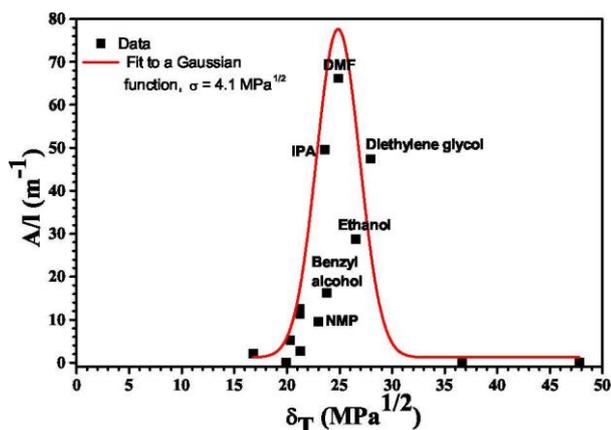

Figure 3: A/l values of MOFN/solvent dispersions as function of solvent Hildebrand solubility parameter.

Figure 3 shows the A/l values of MOFN/solvent dispersions as a function of the Hilde-brand solubility parameter $S_T$. The most successful solvent was N,N-Dimethylformamide (DMF) with $S_T = 24.8$ MPa$^{1/2}$. Other successful solvents include isopropanol (IPA), di- ethylene glycol, and ethanol, in descending order according to A/l values, with Hildebrand solubility parameter in the range $23.5 < S_T < 27.9$ MPa$^{1/2}$.

It is important to notice that the A/l versus $S_T$ data approximately followed a Gaussian function (red solid line in Figure 3) where the most successful solvents were around the mean and the less successful solvents were localized around the tails. This implies that the Hildebrand solubility parameter was able to discriminate between good solvents and less successful solvents. An exception to this conclusion was benzyl alcohol and N-methyl-2- pyrrolidone (NMP).

A similar analysis was conducted with respect to Hansen solubility parameters. Figure S2 (SI, section S3.1) shows A/l values of MOFN/solvent dispersions as a function of each Hansen solubility parameter, i.e. $S_D$, $S_P$, and $S_H$. As expected from equation S8 and in accordance with the previous analysis based on the Hildebrand solubility parameter, the three Hansen solubility parameters identified the same most successful solvents for exfoliation: DMF, IPA, diethylene glycol and ethanol, in that order. The successful solvents had $15.8$ MPa$^{1/2} < S_D < 17.4$ MPa$^{1/2}$, $6.1$ MPa$^{1/2} < S_P < 13.7$ MPa$^{1/2}$ and $11.3$ MPa$^{1/2} < S_H < 19.4$ MPa$^{1/2}$. Figure S2a shows clearly that the dispersive bonding component $S_D$ had a well-defined value within a narrow range for successful solvents indicating that the exfoliation of MOFN had a strong dependence on London dispersive forces. The $S_D$ parameter, however, did not discriminate successful solvents from unsuccessful solvents. The other Hansen solubility parameters presented a more scattered data and further information about the dependence of the exfoliation of MOFN on polar or hydrogen bonding could not be inferred.

The analysis of A/l values versus solvent surface energy/surface tension (SI section S3.2) showed that the surface energy was unable to discriminate successful from unsuccessful solvents for exfoliation of MOFN. For the case of DMF, determined to be the best solvent for exfoliation according to the Hildebrand solubility parameter, the solvent surface energy was indeed very close to the MOFN surface energy. According to the Equation S3 (SI sections S2 and S3.2), the surface energy of DMF is 64.4 mJ m$^{-2}$. The surface energy of the MOFN was determined by the contact angles measurement method (SI section S4) as 60.95 mJ m$^{-2}$. Interestingly, the obtained value is within the range of surface energies measured for other materials (70-80 mJ m$^{-2}$) including graphite[46] and metal chalcogenides.[6]

3.2.2 Optimization of the liquid-phase exfoliation processing parameters

Although DMF was found to be the most successful solvent to exfoliate MOFN, IPA was kept as the preferred solvent due to clear advantages over DMF in terms of toxicity and, as pointed out before, easier processability of obtained dispersions.

The yield of dispersions strongly depends on the exfoliation processing parameters.[6,39] Optimization of the liquid-phase exfoliation process was carried out observing four key parameters: ultrasonication frequency, ultrasonication power, initial concentration of the MOFN/IPA mixture and processing time (SI section S5). The criterion for selection of exfoliation processing parameters was the concentration of the resulting dispersions determined using UV-Vis spectroscopy. Corresponding A/l values (at 550 nm) were plotted as a function of each processing parameter as shown in Figure 4.

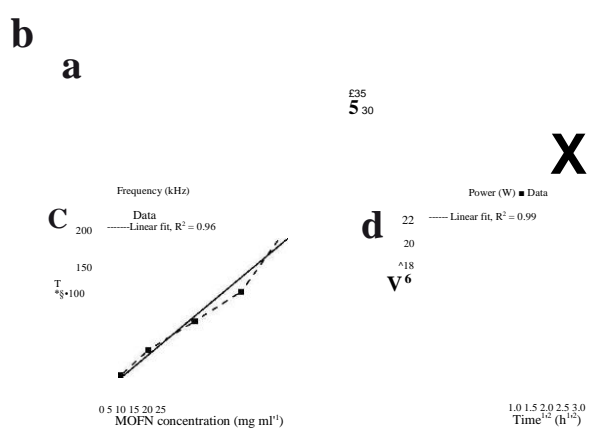

Figure 4: A/l values of MON/IPA dispersion as function of (a) ultrasonication frequency, (b) ultrasonication power, (c) initial concentration of the MOFN/IPA mixture, and (d) square root of the processing time.

It is clear from Figure 4 that optimum processing parameters for the exfoliation of MOFN in IPA to produce dispersions of maximum concentrations were: initial concentration of MOFN/IPA mixture of 20 mg ml$^{-1}$, ultrasonication power of 99 W (30 % of a nominal power of 330 W) and ultrasonication frequency of 80 kHz. Figure 4d shows that the concentration of the MON/IPA dispersion scales linearly with the square root of the processing time. The same dependance of concentration as a function of processing time was reported for the case of exfoliation of graphite to produce graphene.[47] The concentration of a MON/IPA dispersions produced using the optimized processing parameters and processed for 9 hours was 0.1 mg ml$^{-1}$ and the extinction coefficient was $a = 904.08$ m$^{-1}$ L g$^{-1}$ (SI section S6).

3.2.3 MON synthesis mechanism

The starting material MOFN has been synthesized by a typical co-precipitation method where Mn+$^2$ ions are oxidized by KMnO$_4$ in the presence of $EO_{20}PO_{70}EO_{20}$ triblock copolymer *Pluronic* P123.[41] The mechanism or formation of the products is not well known in the literature but few authors have proposed likely mechanisms. Briefly, it is believed that the Mn+$^2$ ions first coordinate with the surfactant P123 and then get oxidized by KMnO$_4$ giving place to a primary structure of MnO$_2$ formed of edge-sharing MnO$_6$ octahedra, next these structures self-assemble into nanosheets possibly having inter-layer K+ cations and H$_2$O molecules.[48,49] Finally the nanosheets assemble into the flower-like nanostructures to minimize surface energy.[41,50,51] It is known that when the synthesis is carried out at room temperature, the product is typically an amorphous material and the crystallinity increases only when using higher synthesis temperatures under hydrothermal conditions.[41,48] In this work, the product was basically a poorly crystalline flower-like nanostructure formed of nanosheets (MOFN) as shown in the images in Figures 1a to 1c and the XRD spectrum in Figure 2.

Upon ultrasonication of the MOFN/IPA mixture, several phenomena affected the MOFN nanostructure. First, we speculate that there was a process of scission/cleavage of the MOFN

"petals" to give place to the MON.[4,52-54] As shown in Figures S10a to S10c, during ultrason- ication numerous MON were produced and typically lay around MOFN cores that appeared to be "unraveled". This view is supported by the fact that the typical size of the resulting MON was 20-40 nm, approximately the same size of the "petals" of the MOFN shown in Figure 1c. We attempted the search of "crystal" defects as points of initiation of the cleavage process using TEM. However, this was not possible due to the poorly crystalline nature of MOFN. Second, we speculate that there was a process of exfoliation of the MON after or perhaps simultaneous with the cleavage from the MOFN core. Due to the poorly crystalline nature of the sample, this exfoliation process must have been substantially different from that undergone by crystalline van der Waals solids[2] and therefore a more appropriate term for it is an "exfoliation-like" process. Two pieces of evidence support the occurrence of such a process. First, the thickness of several MOFN "petals" was measured before any processing using the *log-ratio technique,* and the average thickness determined over 10 "petals" was 8.2 ± 1.9 nm. Therefore, it is possible that the MOFN "petals" had undergone an exfoliationlike process giving place to at least two nanosheets (average thickness 3.2 ± 1.2 nm). Second, the steps observed at the edges of some MON (Figures S10d to S10f) likely resulted from the ongoing exfoliation-like process. We can conclude that the MON were likely produced mostly by scission and to a lesser extent by an exfoliation-like process. For the sake of simplicity, we chose to keep the name of "exfoliation" to refer to the overall process giving place to MON.

## 3.3 Synthesis and characterization of graphene-manganese oxide nanosheets hybrid

The design of hybrid-2D-nanomaterials of superior properties by strategically combining two or more 2D nanomaterials holding key complementary properties is of high relevance in many application areas. The synthesis and application of hybrid-2D-nanomaterials implies major challenges including: (1) compatibility of synthesis procedures followed for each 2D-nanomaterial, (2) stability of the hybrid suspension against re-aggregation, (3) achieve ment of a desired interface between the 2D nanomaterials, (4) tuning of size, thickness, shape and crystallographic orientation of one 2D nanomaterial in relation to another, (5) reproducibility and scalability of synthesis procedures and (6) compatibility of the hybrid synthesis procedure with further processing steps in the manufacture of devices.

In this study, we achieved for the first time the co-exfoliation of graphite and MOFN in IPA using a single-step process, which resulted in a hybrid of graphene and manganese oxide nanosheets (GMOH). The primary application pursued for this 2D-hybrid was energy storage where the role of graphene is to provide electrical conductivity while the role of the manganese oxide nanosheets is to provide pseudocapacitive activity. The GMOH showed an enhanced performance for supercapacitor applications as described somewhere else (a work currently in preparation).

Suspensions of GMOH were obtained by a single-step co-exfoliation in IPA of graphite and MOFN following the processing parameters described in the subsection '*Experimental methods/Synthesis of GMOH dispersion*'). Figures 5a-5c show TEM images of the GMOH exfoliated for 3, 9 and 18 hours. Upon drop-casting onto TEM grids, the manganese oxide nanosheets (20-40 nm lateral dimension) lay on top of the larger graphene flakes (400 nm lateral dimension and thickness ~ 6.5 ± 2.4 nm, as determined by the *log-ratio technique* described in the SI section S7). For a processing time of 18 hours, the GMOH dispersion was largely more concentrated and as shown in Figure 5c, the MON appeared to be more crowded and had an evident degree of aggregation.

Raman spectroscopy was used to confirm the simultaneous presence of graphene and MON in the GMOH. Samples of MON and graphene exfoliated separately in IPA were used for reference. Figure 6 shows the Raman spectra of MON, GMOH and graphene respectively. Both graphene and GMOH showed characteristic G (1345.9 cm$^{-1}$, 1352.6 cm$^{-1}$), D (1563.7 cm$^{-1}$, 1580.4 cm-1) and 2D bands (2681.2 cm$^{-1}$, and 2702.3 cm$^{-1}$),[39,40] confirming the presence of graphene in both the as-exfoliated graphite and the GMOH sample. MON showed a peak at 629.2 cm$^{-1}$ and a broad peak at about 296.2 cm$^{-1}$, which were present also in GMOH at 655.6 cm$^{-1}$ and 335.4 cm$^{-1}$. Gao T. *et al.* associated the presence of a peak at 634.0 cm$^{-1}$ with $A_g$ spectroscopic modes that originate from the breathing vibrations of the MnO$_6$ octahedra present within tetragonal hollandite-type framework in a-MnO$_2$.[43,55] Moreover, peaks at 510 cm$^{-1}$ and 580 cm$^{-1}$ have been reported as characteristic of a-MnO$_2$; it is however also well known that these peaks tend to disappear rather quickly as a-MnO$_2$ very easily degrades under laser-induced heating giving place to new phases[56,57] generating peaks at 310 cm$^{-1}$, 360-390 cm$^{-1}$, 633 cm$^{-1}$, 642 cm$^{-1}$ (all attributed to the presence of Mn$_2$O$_3$[56,57]), and 310 cm$^{-1}$, 315 cm$^{-1}$, and 650 cm$^{-1}$ (all attributed to the presence of Mn$_3$O$_4$[57]). This confirms that the poorly crystalline a-MnO$_2$ precursor was degraded into Mn$_2$O$_3$ and Mn$_3$O$_4$ due to laser-induced heating during the acquisition of Raman spectra. We nevertheless proved that the GMOH effectively incorporated simultaneously both exfoliated graphene and MON species. This conclusion was supported by TGA and XPS measurements (SI sections S8 and S9).

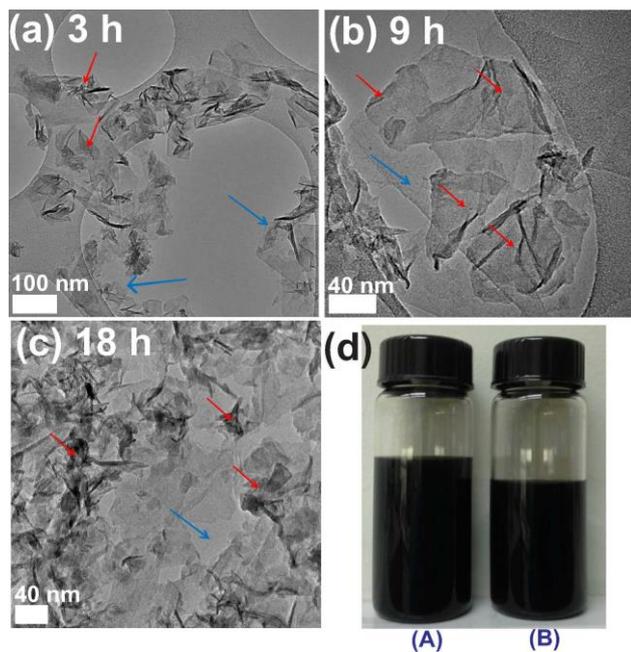

Figure 5: TEM images of GMOH exfoliated for (a) 3 hours, (b) 9 hours, (c) 18 hours, and (d) vials containing GMOH/IPA dispersions exfoliated for 9 hours:(A) as-prepared and (B) stable after 6 months. The arrows point MON (red) and graphene (blue).

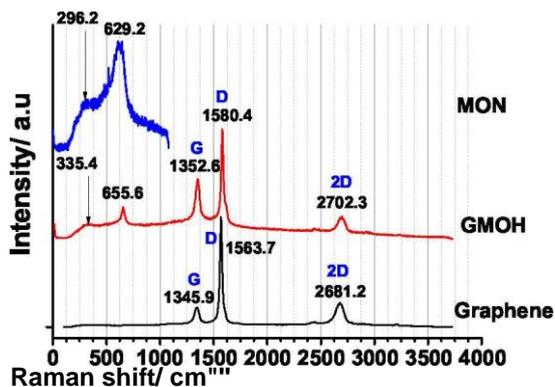

Figure 6: Raman spectra of graphene, MON and GMOH. Numeric labels identify peak position (cm$^{-1}$) and text labels identify characteristic bands of graphene.

3.3.1 Stability of GMOH suspensions

The stability of GMOH/IPA (processed for 9 hours) dispersions was assessed by Z potential measurements. For an overall evaluation, the single component MON/IPA and graphene/IPA dispersions were also considered. All dispersions had a pH =7. Figure 7 shows the Z-potential curves and the Table 1 summarizes the average Z-potentials of all dispersions. A Z potential > + 25 mV for all the dispersions warranted stability against aggregation. [58] All the dispersions had a positive Z potential increasing in the order MON/IPA < graphene/IPA < GMOH/IPA. These results indicated that the positive Z potentials of graphene and MON in IPA causes a repulsive effect that stabilizes the GMOH/IPA dispersion, hence having an overall larger positive Z potential. The reason for the graphene and MON species to have positive Z potentials in IPA, however, is not clear. The surface chemistry at the interface between the solute/solvent should be further investigated in future work. Stable GMOH/IPA dispersions after 6 months are shown in Figure 5d.

Table 1: Average Z-potentials (over 3 acquired measurements) for MON/IPA, graphene/IPA and GMOH/IPA.

| Sample | Z-potential (mV) |
|---|---|
| MON/IPA | 26.8 ± 5.5 |
| Graphene/IPA | 34.8 ± 2.8 |
| GMOH/IPA | 48.1 ± 4.5 |

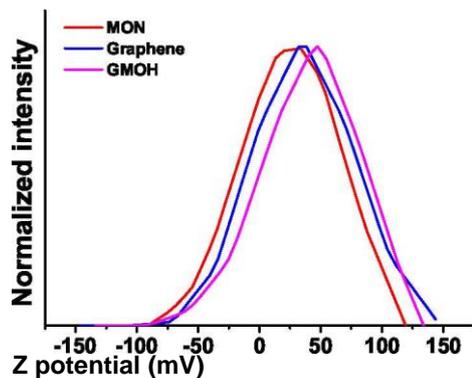

Figure 7: Average Z-potential curves (over 3 acquired measurements ) for MON/IPA, graphene/IPA and GMOH/IPA dispersions.

3.3.2 The co-exfoliation process

This section provides some experimental facts that shed light on the co-exfoliation process. In order for the simultaneous exfoliation of two solutes in a single solvent to occur and produce a stable suspension: (1) the overall process must be energetically favorable, (2) the solutes must have common solubility properties, and (3) the interaction of exfoliated solutes must comply with conditions that avoid post-exfoliation aggregation.

Examination of Hildebrand and Hansen solubility parameters of solvents that best exfoliate MOFN to obtain MON (studies in this work) and graphite to obtain graphene (obtained from the literature[59]) showed similar $\delta_D$, $Sp$ solubility parameters(detailed in the SI). This indicated that MOFN and graphite have similar solubility properties. In the case of IPA, in terms of concentration of the resulting dispersions, this solvent is more suitable to exfoliate MOFN than graphite. Therefore, a larger proportion of MON than graphene was expected during co-exfoliation in IPA. This was indeed verified experimentally as described next.

Using identical processing parameters, described in the subsection *'Experimental methods/Dispersions for co-exfoliation studies',* MON/IPA, graphene/IPA and GMOH/IPA dispersions were synthesized for 3, 6, 9, 15 and 18 hours and their concentrations were determined. The concentration of the solutes of the starting mixtures MOFN/IPA and graphite/IPA was 10 mg ml$^{-1}$. In the case of the starting mixture graphite/MOFN/IPA, the concentrations of MOFN and graphite were 10 mg ml$^{-1}$ each. The overall concentration of the resulting GMOH/IPA dispersion was determined by vacuum filtration, the wt% of MON and graphene were determined using TGA (Table S5) and used to calculate their respective concentrations.

Figure 8 shows a graph of the concentration of the MON/IPA, graphene/IPA and GMOH/IPA dispersions as a function of the square root of the processing time. The curves corresponding to the concentration of MON and graphene in the GMOH are also shown. The GMOH/IPA dispersions showed larger concentrations (up to 0.5 mg ml$^{-1}$) than the MON/IPA (up to 0.45 mg ml$^{-1}$) and graphene/IPA (up to 0.02 mg ml$^{-1}$) dispersions at all processing times. In the GMOH, the wt% of MON (85.3 wt% to 77.7 wt%) was larger than the wt% of graphene (14.7 wt% to 22.3 wt%) as predicted by the Hildebrand and Hansen solubility parameters. Importantly, at the same starting concentration of graphite and MOFN (10 mg ml$^{-1}$), the quantities of MON and graphene obtained by the co-exfoliation procedure were larger than the quantities obtained by the single-solute exfoliation procedures. This effect was more pronounced for graphene (up to 4 times more graphene produced by co-exfoliation, 0.09 mg ml$^{-1}$ > 0.02 mg ml$^{-1}$) than for MON (up to 1.1 times more MON produced by co-exfoliation, 0.2 mg ml$^{-1}$ > 0.18 mg ml$^{-1}$). This implies that the presence of a second solute, MOFN, enhances the exfoliation of graphite at all processing times. A study of the thermodynamics involved in the co-exfoliation should suitably complement these experimental findings. We can conclude that our experimental findings indicated that the co-exfoliation procedure here studied is energetically favorable and produces stable suspensions of enhanced yield as compared to the single-solute exfoliation procedure.

The advantages of the synthesis of the GMOH/IPA dispersion by co-exfoliation can be summarized as follows: (1) it is a facile single-step procedure where two 2D-nanomaterials were synthesized simultaneously, (2) the resulting GMOH/IPA dispersion was stable enabling further processing steps, (3) upon drop-casting or spray depositing, the manganese oxide nanosheets deposited directly onto the graphene flakes, a desirable interaction achieved without the need for surface chemistry modification. Previously, we had followed a different approach where each solute was exfoliated in a different solvent followed by mixing of the resulting dispersions. Aggregation of the solutes was frequently encountered hampering any further processing. The production of active materials in suspensions implies that the co-exfoliation procedure is compatible with other techniques such as spray-deposition and ink-jet printing for the manufacture of films. The co-exfoliation procedure can be extended to other solutes provided that a solvent with suitable solubility parameters for two or more solutes are carefully chosen.

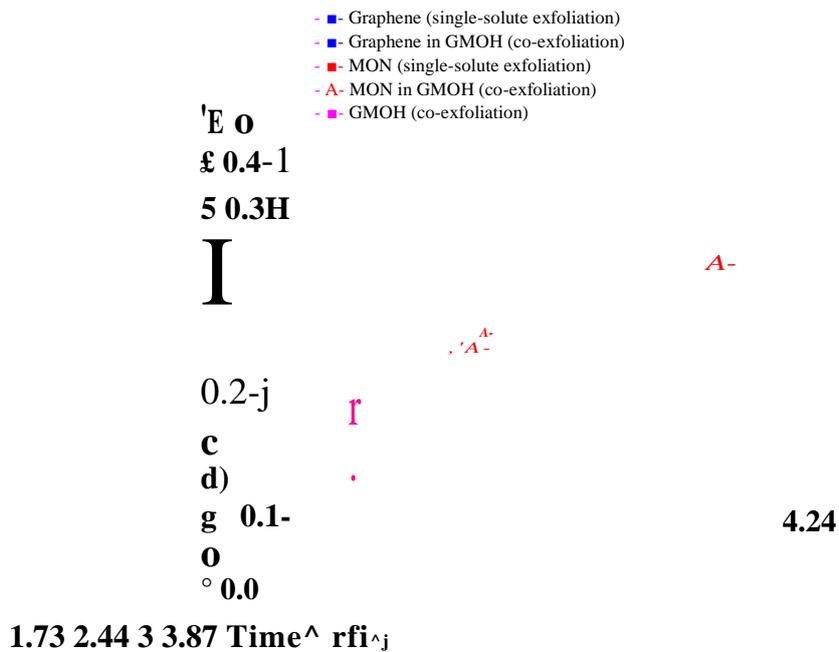

Figure 8: Concentration of suspensions prepared by single-solute exfoliation (MON/IPA and graphene/IPA) and co-exfoliation (GMOH/IPA) as a function of the square root of processing time. The concentration of MON and graphene in the GMOH are also shown.

# 4 Conclusions

We have demonstrated the synthesis of manganese oxide nanosheets by a liquid-phase exfoliation procedure. The manganese oxide nanosheets were produced in a stable suspension in IPA with concentrations up to 0.45 mg ml$^{-1}$. We demonstrated the co-exfoliation of graphite and manganese oxide with a flower-like nanostructure to give place to stable suspensions of a 2D hybrid of graphene and manganese oxide nanosheets with concentrations up to 0.5 mg ml$^{-1}$. The co-exfoliation procedure here followed was found to be energetically favorable producing stable two-solutes suspensions where the yield of each solute was superior to that obtained by a single-solute exfoliation procedure. This work opens up new venues of research focused on the investigation of co-exfoliation of other organic and inorganic 2D materials highly desirable for many technological applications.

## Acknowledgement

The authors thank Niall McEvoy and Prof Georg Duesberg (CRANN,TCD) for their help with the Raman characterization, Philip Holdway and Alison Crossley (Oxford Materials Characterization Services) for their help with XRD and XPS characterization, as well as the Advanced Microscopy Laboratory and CRANN for access to their facilities. The authors wish to acknowledge support from the European Research Council (ERC Starting Grant 2DNanoCaps) and Science Foundation Ireland.

## Supporting Information Available

Synthesis of MOFN at various temperatures; Theoretical background of liquid-phase exfoliation; Solvents selection including Hansen solubility parameters and surface energy and surface tension; Determination of surface energy of MOFN; Optimization of the liquid-phase exfoliation processing parameters; Determination of extinction coefficient of the MON/IPA dispersion; Determination of thickness of MON and graphene flakes; Thermogravimetric analysis; XPS analysis; Analysis of the solubility parameters of MON and graphene; and MON synthesis mechanism.

This material is available free of charge via the Internet at `http://pubs.acs.org/`.

References

(1) Novoselov, K. S.; Geim, A. K.; Morozov, S. V.; Jiang, D.; Zhang, Y.; Dubonos, S. V.; Grigorieva, I. V.; Firsov, A. A. *Science* 2004, *306,* 666-669.

(2) Butler, S. Z. et al. *ACS Nano* 2013, 7, 2898-2926.

(3) Mas-Balleste, R.; Gomez-Navarro, C.; Gomez-Herrero, J.; Zamora, F. *Nanoscale* 2011, 3, 20-30.

(4) Chhowalla, M.; Shin, H. S.; Eda, G.; Li, L.-J.; Loh, K. P.; Zhang, H. *Nature Chem.* 2013, 5, 263-275.

(5) Nicolosi, V.; Chhowalla, M.; Kanatzidis, M. G.; Strano, M. S.; Coleman, J. N. *Science* 2013, *340*.

(6) Coleman, J. N. et al. *Science* 2011, *331,* 568-571.

(7) Smith, R. J.; King, P. J.; Lotya, M.; Wirtz, C.; Khan, U.; De, S.; O'Neill, A.; Dues- berg, G. S.; Grunlan, J. C.; Moriarty, G.; Chen, J.; Wang, J. Z.; Minett, A. I.; Nicolosi, V.; Coleman, J. N. *Adv. Mater.* 2011, *23,* 3944.

(8) Ma, R.; Sasaki, T. *Adv. Mater.* 2010, 22, 5082-5104.

(9) Bizeto, M. A.; Shiguihara, A. L.; Constantino, V. R. L. *J. Mater. Chem.* 2009, *19,* 2512-2525.

(10) Lukatskaya, M. R.; Mashtalir, O.; Ren, C. E.; DallAgnese, Y.; Rozier, P.; Taberna, P. L.; Naguib, M.; Simon, P.; Barsoum, M. W.; Gogotsi, Y. *Science* 20 1 3 , *341,* 1502-1505.

(11) Mak, K. F.; Lee, C.; Hone, J.; Shan, J.; Heinz, T. F. *Phys. Rev. Lett.* 2010, *105,* 136805, PRL.

(12) Osada, M.; Sasaki, T. In *Nanofabrication*; Masuda, Y., Ed.; Intech, 2011; Chapter 5.

13  Osada, M.; Sasaki, T. *Adv. Mater.* 2012, *24,* 210-228.

14  L., W.; Takada, K.; Kajiyama, A.; Onoda, M.; Michiue, Y.; Zhang, L.; Watanabe, M.; Sasaki, T. *Chem. Mater.* 2003, 15, 4508-4514.

15  Suzuki, S.; Miyayama, M. *J. Phys. Chem. B* 2006, *110,* 4731-4734.

16  Xin, H.; Ma, R.; Wang, L.; Ebina, Y.; Takada, K.; Sasaki, T. *Appl. Phys. Lett.* 2004, *85,* 4187-4189.

Miyamoto, N.; Kuroda, K. *J Colloid Interf. Sci.* 2007, *313,* 369-373.

17  Sasaki, T.; Nakano, S.; Yamauchi, S.; Watanabe, M. *Chem. Mater.* 1997, *9,* 602-608.

18  Iida, M.; Sasaki, T.; Watanabe, M. *Chem. Mater.* 1998, *10,* 3780-3782.

19  Zhao, X.; Mendoza-Sanchez, B.; Dobson, P. J.; Grant, P. S. *Nanoscale* 2011, 3, 839855.

20  Xu, C.; Kang, F.; Li, B.; Du, H. *J. Mater. Res.* 2010, *25,* 1421-1432.

Wei, W.; Cui, X.; Chen, W.; Ivey, D. G. *Chem. Soc. Rev.* 2011, 40, 1697-1721.

21  Toupin, M.; Brousse, T.; Beianger, D. *Chem. Mater.* 2004, *16,* 3184-3190.

22  Ge, H.; Yu, R.; Mi, D.; Zhu, Y. Removal of $NO_2$ and $O_3$ generated from corona discharge in indoor air cleaning with $MnO_2$ catalyst. Journal of Physics: Conference Series. 2013; p 012120.

24  Saputra, E.; Muhammad, S.; Sun, H.; Ang, H. M.; Tade, M. O.; Wang, S. *Environ. Sci. Tech.* 2013, *47.*

27) Fei, J.; Cui, Y.; Yan, X.; Qi, W.; Yang, Y.; Wang, K.; He, Q.; Li, J. *Adv. Mater.* 2008, *20,* 452-456.

25  Dong, W. F.; Zang, L. H.; Li, H. *Appl. Mech. Mater.* 2013, *361,* 760-763.

(28) Sakai, N.; Ebina, Y.; Takada, K.; Sasaki, T. *J. Electrochem. Soc.* 2005, 152, E384- E389.

(29) Omomo, Y.; Sasaki, T.; Wang, L.; Watanabe, M. *J. Am. Chem. Soc.* 2003, 125, 35683575, doi: 10.1021/ja021364p.

(30) Liu, Z.-h.; Ooi, K.; Kanoh, H.; Tang, W.-p.; Tomida, T. *Langmuir* 2000, *16,* 4154-4164.

(31) Xu, C.; Shi, S.; Sun, Y.; Chen, Y.; Kang, F. *Chem. Commun.* 2013,

(32) Shi, S.; Xu, C.; Yang, C.; Chen, Y.; Liu, J.; Kang, F. *Sci. Rep.* 2013, *3*.

(33) Zhang, J.; Jiang, J.; Zhao, X. S. *J. Phys. Chem. C* 2011, *115,* 6448-6454.

(34) Kai, K.; Yoshida, Y.; Kageyama, H.; Saito, G.; Ishigaki, T.; Furukawa, Y.; Kawa-mata, J. *J. Am. Chem. Soc.* 2008, *130,* 15938-15943.

(35) Zhao, G.; Li, J.; Jiang, L.; Dong, H.; Wang, X.; Hu, W. *Chem. Sci.* 2012, 3, 433-437.

(36) Wei, L.; Li, C.; Chu, H.; Li, Y. *Dalton T.* 2011, *40,* 2332-2337.

(37) Liu, J.; Jiang, J.; Cheng, C.; Li, H.; Zhang, J.; Gong, H.; Fan, H. J. *Adv. Mater.* 2011, *23,* 2076-2081.

(38) Li, W.; Li, G.; Sun, J.; Zou, R.; Xu, K.; Sun, Y.; Chen, Z.; Yang, J.; Hu, J. *Nanoscale* 2013, *5,* 2901-2908.

(39) Hernandez, Y. et al. *Nat. Nano* 2008, 3, 563-568.

(40) Mendoza Sanchez, B.; Rasche, B.; Nicolosi, V.; Grant, P. *Carbon* 2012, *52,* 337-346.

(41) Jiang, H.; Sun, T.; Li, C.; Ma, J. *J. Mater. Chem.* 2012, 22, 2751-2756.

(42) Wang, J.; Khoo, E.; Ma, J.; See Lee, P. *Chem. Commun.* 2010, 46, 2468-2470.

(43) Julien, C. M.; Massot, M.; Poinsignon, C. *Spectrochim. Acta A Mol. Biomol. Spectrosc.* 2004, *60,* 689-700.

(44) Brousse, T.; Toupin, M.; Dugas, R.; Athoul, L.; Crosnier, O.; Blanger, D. *J. Elec- trochem. Soc.* 2006, *153,* A2171-A2180.

(45) Long, J. W.; Swider-Lyons, K. E.; Stroud, R. M.; Rolison, D. R. *Electrochem. Solid St.* 2000, 3, 453-456.

(46) Lotya, M.; Hernandez, Y.; King, P. J.; Smith, R. J.; Nicolosi, V.; Karlsson, L. S.; Blighe, F. M.; De, S.; Wang, Z.; McGovern, I. T.; Duesberg, G. S.; Coleman, J. N. *J. Am. Chem. Soc.* 2009, *131,* 3611-3620.

(47) Khan, U.; O'Neill, A.; Lotya, M.; De, S.; Coleman, J. N. *Small* 2010, *6,* 864-871.

(48) Subramanian, V.; Zhu, H.; Vajtai, R.; Ajayan, P. M.; Wei, B. *J Phys. Chem. B* 2005, *109,* 20207-20214.

(49) Yan, D.; Yan, P. X.; Yue, G. H.; Liu, J. Z.; Chang, J. B.; Yang, Q.; Qu, D. M.; Geng, Z. R.; Chen, J. T.; Zhang, G. A.; Zhuo, R. F. *Chem. Phys. Lett.* 2007, *440,* 134-138.

(50) Min, Y.; Akbulut, M.; Kristiansen, K.; Golan, Y.; Israelachvili, J. *Nat. Mater.* 2008, *7,* 527-538.

(51) Gao, Y. J.; Zhang, W. C.; Wu, X. L.; Xia, Y.; Huang, G. S.; Xu, L. L.; Shen, J. C.; Siu, G. G.; Chu, P. K. *Appl. Surf. Sci.* 2008, *255,* 1982-1987.

(52) Los, S.; Duclaux, L.; Alvarez, L.; Hawelek, L.; Duber, S.; Kempinski, W. *Carbon* 2013, *55,* 53-61.

(53) Wang, W.; Wang, Y.; Gao, Y.; Zhao, Y. *J. Supercrit. Fluid* 2014, *85,* 95-101.

(54) Junggeburth, S. C.; Diehl, L.; Werner, S.; Duppel, V.; Sigle, W.; Lotsch, B. V. *J. Am. Chem. Soc.* 2013, *135,* 6157-6164.

(55) Gao, T.; Fjellvg, H.; Norby, P. *Anal. Chim. Acta* 2009, *648,* 235-239.

(56) Luo, J.; Zhu, H. T.; Fan, H. M.; Liang, J. K.; Shi, H. L.; Rao, G. H.; Li, J. B.; Du, Z. M.; Shen, Z. X. *J. Phys. Chem. C* 2008, 112, 12594-12598.

(57) Buciuman, F.; Patcas, F.; Craciun, R.; Zahn, R. D. *Phys. Chem. Chem. Phys.* 1999, 1, 185-190.

(58) Bogdan, N.; Vetrone, F.; Ozin, G. A.; Capobianco, J. A. *Nano Lett.* 2011, 11, 835-840.

(59) Bergin, S. D.; Sun, Z.; Rickard, D.; Streich, P. V.; Hamilton, J. P.; Coleman, J. N. *ACS Nano* 2009, 3, 2340-2350.

Graphical TOC Entry

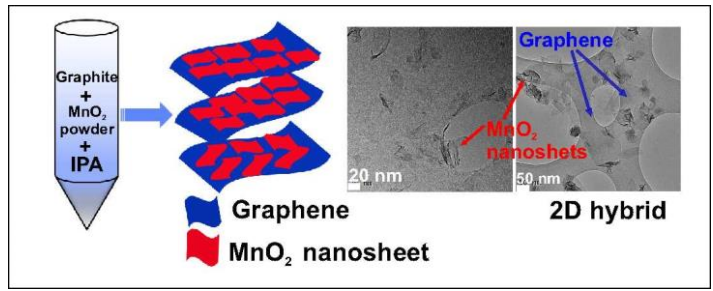